# A Table-Top Formation of Bilayer Quasi-Free-Standing Epitaxial-Graphene on SiC(0001) by Microwave Annealing in Air

Kwan-Soo Kim[1], Goon-Ho Park[1], Hirokazu Fukidome[1], Someya Takashi[2], Iimori Takushi[2], Komori Fumio[2], Matsuda Iwao[2], and Maki Suemitsu[1]

[1]Research Institute of Electrical Communication, Tohoku University, Sendai, 980-8577, Japan
[2]Institute for Solid State Physics, The University of Tokyo, Chiba 277-8581, Japan

We propose a table-top method to obtain bilayer quasi-free-standing epitaxial-graphene (QFSEG) on SiC(0001). By applying a microwave annealing in air to a monolayer epitaxial graphene (EG) grown on SiC(0001), the $(6\sqrt{3} \times 6\sqrt{3})R30°$ reconstructed buffer layer is decoupled from the SiC substrate and becomes the second EG layer as confirmed by the low energy electron diffraction, high-resolution transmission electron microscopy, Raman scattering spectroscopy, X-ray photoelectron spectroscopy, and angle-resolved photoelectron spectroscopy. The most likely mechanism of the decoupling is given by the oxidation of the SiC surface, which is quite similar to what happens in conventional annealing method in air but with a process time by more than one order of magnitude less.

1. Introduction

Graphene, a single layer of $sp^2$-hybridized carbon atoms, is one of the most advanced two-dimensional materials.[1-3] Because of its unique electronic, optical, mechanical and thermal properties, graphene is considered as one of the promising candidate materials for future nano-scaled electronic devices. Among various fabrication methods, epitaxial growth of graphene on a single crystal SiC substrate using thermal decomposition process is industrially quite practical since it provides a high-quality graphene with a large-scale, transfer-free manner.[4-6] For the epitaxial graphene (EG) growth on the Si-terminated SiC(0001), the $(6\sqrt{3} \times 6\sqrt{3})R30°$ reconstructed buffer layer is formed between graphene and the SiC substrate. The buffer layer has a graphene-like honeycomb structure but with covalent bonds between its C atoms and the Si atoms in the SiC surface, and is thus electrically inactive.[7, 8] Moreover, this layer is even electrically harmful in that it causes high electron doping (~$10^{13}$ cm$^{-2}$) in EG and contains phonon scattering centers, which degrades the carrier mobility to ~$10^3$ cm$^2$/Vs.[9-11]

As such, elimination of the buffer layer has been sought to restore the natural properties of graphene. The most typical method is the hydrogen intercalation.[12-14] By annealing EG in a H$_2$ ambient at ~700 $^o$C, we can diffuse hydrogen molecules towards the interface between the buffer layer and the SiC crystal, and break the covalent bondings in between. The buffer layer can then be transformed into a mono-layer of graphene, becoming a so-called quasi-free-standing epitaxial graphene (QFSEG). Consequently, the doping type changes from the original n-type into p-type with the charge density of ~$10^{12}$ cm$^{-2}$ and the mobility is increased by a factor of three.[13, 14] Aside from hydrogen, various intercalation methods have been reported, which includes lithium,[15] germanium,[16] fluorine[17, 18] and gold[19]. In these methods, the formation of QFSEG proceeds via thermal diffusion of these elements into the buffer layer. During this process, however, defects are generated by intercalated atoms and it is difficult to form homogeneous QFSEG over a large area. In this situation, Oliveira et al [20] have shown that a homogeneous QFSEG can be formed by thermal annealing EG in air, in which the oxygen atoms in the air break the covalent Si-C bondings through oxidation of the Si atoms. According to them, QFSEG can be formed without inducing defects in the topmost graphene layer.[20-22] Recently, Bao et al.[23] have succeeded in the synthesis of QFSEG by a rapid cooling after a thermal heating at ~800 $^o$C using a quenching into a liquid nitrogen. In this method, QFSEG is considered to be formed via physical breakage of the covalent bondings because graphene shows a negative thermal expansion coefficient (TEC) as opposed to SiC.

Despite the variety of methods, all the existing intercalation methods utilize a thermal heating process in a furnace, in which the thermal energy is transferred to the material through convection and radiation, and then via diffusion from the surface to the bulk. This indirect nature of the heating mechanism might be the cause for the inhomogeneity observed frequently. Here we propose a totally new heating method, microwave annealing (MWA), in which heat is generated inside the bulk of the material by absorbing electromagnetic power.[24-26] The microwave heating is thus gifted with a high thermal uniformity, rapid heating/cooling rates, selectivity of the material to be heated, and suppression of unexpected diffusion of species. The microwave heating is now being applied to device fabrication processes in the field of organic and metal-oxide electronics.[27-29] The method has been applied to graphene processings as well, such as in chemical reduction of graphene oxide[30] and exfoliation of graphite[31], but with the help of agents in either aqueous or organic media. On direct heating of samples without use of agents, however, the studies have been limited. While Kim *et al.*[32] confirmed onset of hole doping by oxygen in air in exfoliated graphene and Lee *et al.*[33] reported onset of exfoliation and nitrogen doping of graphite, both by microwave irradiation, no studies have ever been made on the QFSEG formation using microwave. Here we will report the first application of MWA in the formation of QFSEG. In hindsight, the SiC substrate used in QFSEG was an ideal material for MWA in that it has a high tangent loss, a measure for the conversion efficiency from the electromagnetic energy to the thermal energy.[34]

2. **Experimental methods**

A 6H-SiC (0001) wafer (II-VI Inc., semi-insulating) was cut into $10\times10$ mm$^2$ pieces, which were cleaned in acetone, ethanol and HF. Monolayer epitaxial graphene (mono-EG) was grown on the substrates by annealing them under an Ar atmosphere at 1325 $^o$C for 20 min using a conventional furnace. MWA was carried out using a home-use microwave oven under atmospheric ambient with a fixed frequency of 2.45 GHz with the output power fixed at 1000 W. In this condition, the temperature of sample easily goes up to 500 $^o$C in 120 s. Raman scattering spectroscopy ($\lambda$=514 nm), low energy electron diffraction (LEED), high-resolution transmission electron microscope (HR-TEM), atomic force microscopy (AFM), X-ray photoelectron spectroscopy (XPS), angle-resolved photoelectron spectroscopy (ARPES) and Hall measurement were used to characterized the structural, chemical and electronic properties of the mono-EG before and after MWA.

3. **Results and discussion**

Figure 1a shows the LEED pattern of the pristine mono-EG. It clearly presents six graphene spots, each of which is surrounded by six smaller superstructure spots associated with the reconstructed $(6\sqrt{3} \times 6\sqrt{3})R30º$ structure, a fingerprint of the carbon buffer layer.[35] After the MWA (Figure 1b), this superstructure smeared out and the major graphene spots become brighter. This suggests that the carbon buffer layer was decoupled from the SiC substrate and became the first layer graphene. This suggestion was more directly confirmed in the HR-TEM images (Figures. 1c and 1d), in which the number of graphene layer increased from one to two. Moreover, the distance between the first carbon layer and the top silicon layer of the SiC substrate increased from 2.7 Å to 4.8 Å after MWA. This increase is related to a change of the bonding nature at the interface as we know the presence of a difference between the buffer/SiC distance (~2.5 Å) and the buffer/graphene distance (~3.5 Å).[8] Since this is understood to be due to the difference in the bonding nature of the interfaces, covalent and van der Waals, a similar change in the bonding nature is suggested to be operative in our MWA treatment. We thus conclude that MWA in air can convert the $(6\sqrt{3} \times 6\sqrt{3})R30º$ reconstructed carbon buffer layer to a decoupled graphene monolayer.

To chemically corroborate the effect of MWA, we conducted the X-ray photoelectron spectroscopy (XPS) analysis. The C1s core level spectrum of the pristine graphene can be deconvoluted into three components (Figure 2a). The components centered at 283.66 cm$^{-1}$ (blue line) and 284.71 cm$^{-1}$ (green line) are attributed to the SiC substrate and the graphene, respectively. The components S1 (285.08 eV) and S2 (285.55 eV) are associated with the buffer layer. The lower energy S1 results from the carbon atoms in the buffer layer covalently bonded to the Si atom in the SiC surface while the higher energy S2 is due to the sp$^2$ bonded carbon atoms in the buffer layer.[7] After the MWA in air (Figure 2b), the buffer-layer-related S1 and S2 components disappeared, and only the graphene and the SiC components survived. The binding energy of the C1s SiC component decreased from 283.66 eV to 282.91 eV by 0.75 eV. A similar binding energy shift was observed in the Si2p core level (Figures 3a and 3b), whose SiC component decreased from 100.45 eV to 99.71 eV by 0.74 eV. These shifts toward lower binding energies indicate increase of the upward bending of the SiC bands. In the Si2p spectrum of the pristine mono-EG (Figure 3a), the peak at 101.08 eV indicated as Si-C is attributed to those Si atoms in the topmost surface of the SiC substrate that are bonded to the carbon atoms in the buffer layer. After the MWA, this Si-C component disappears and a pair of new components appeared at 100.40 eV ($Si^+$) and 101.58 eV ($Si^{4+}$). These components are most reasonably attributed to Si atoms bonded to one ($Si^+$) and four ($Si^{4+}$) oxygen atoms.[20,36] These results indicate that the Si atoms in the SiC surface has been oxidized during MWA in air, most probably by absorbing oxygen precursors such as $O_2$ and $H_2O$ embedded in air.

Raman scattering spectroscopy indicated that the quality of the QFSEG is of highest quality. The spectrum of the pristine mono-EG (black line in Figure 4a) consists of a pair of features: the G-band (1596 cm$^{-1}$) and the G'-band (2717 cm$^{-1}$). The broad bands at around 1350 cm$^{-1}$ and 1550 cm$^{-1}$ are attributed to defects (D-band) and to the buffer layer (BL-band), respectively. After the MWA (red line), both the D-band and the BL-band completely disappeared, indicating decoupling of the buffer layer from the SiC substrate without degradation of the graphene quality. In the case of the hydrogen intercalation, it is reported that the intercalated hydrogen atoms passivate the dangling bonds at the grain boundaries and thereby improve the quality of graphene.[14] Although a similar betterment mechanism is expected in the case of oxygen intercalation, a more detailed analysis is left to future research.

The decoupling of the buffer layer is also confirmed by the width of the G'-band peak. Figure 4b shows the micro-Raman mapping (20×20 μm$^2$) of the full width at half maximum (FWHM) at the G'-band (top) and its histogram (bottom). The FWHM of the G'-band reportedly has a good correlation with the number of graphene layers,[37,38] and the small FWHM value (~33 cm$^{-1}$) of the pristine sample, as shown in the histogram, indicates that it is of monolayer graphene having a good uniformity. After the MWA, the FWHM increased to ~65 cm$^{-1}$, indicating a conversion of the monolayer to bilayer graphene. A noticeable red shift of G'-band position from 2717 cm$^{-1}$ to 2703 cm$^{-1}$ was also observed. The position of G'-band is sensitive to the change of strain in the graphene layer,[39] and the relative change in the lattice constant can be calculated using the expression[13] $\Delta a/a_0 = -(\Delta_{G',\text{strain}}/11.3 \text{ cm}^{-1}) \times 10^{-3}$. Here $\Delta a$ is the relative change of the lattice constant from its equilibrium value $a_0$, and $\Delta_{G',\text{strain}}$ is the shift of the G'-band peak. The black histogram in Figure 4c for the pristine mono-EG (black) has a peak at $\Delta a/a_0 = -3.5 \times 10^{-3}$, indicating onset of a compressive strain. This compression is due to the substantial lattice mismatch between graphene and SiC as well as to a large difference in TEC.[40,41] After the MWA, this compressive strain was relaxed to $\Delta a/a_0 = -2.1 \times 10^{-3}$, which is related to the conversion of the strong -bonding to the weak van der Waals bonding at the graphene/SiC interface.

The results described above clearly indicate that the MWA somehow induces oxygen intercalation at the graphene/SiC interface. This intercalation should bring about a substantial change in the electronic band structure. Figure 5a shows the ARPES analysis, collected at around K point of the graphene Brillion zone. The pristine mono-EG (left) shows a single π-band crossing the Dirac point at an energy of ~0.4 eV below the Fermi level, indicating an electron doping of ~$1\times10^{13}$ cm$^{-2}$.[42, 43] A similar electron concentration (0.8 ~ $1.1\times10^{13}$ cm$^{-2}$) was obtained in the Hall measurement as well (Figure 5b) together with the mobility of 1250 ~ 1750 cm$^2$/Vs. This n-type doping is understood to be due to the buffer layer.[44] After the MWA in air, the π-band splits into two, indicating a formation of the bilayer QFSEG. The Dirac point now sits above the Fermi Level by ~0.16 eV, indicating onset of inversion of the doping type. This upper shift of the Dirac point is qualitatively consistent with the decrease of the binding energy of the C1s graphene peak (Figures 2a). The quantitative difference i.e., 0.56 eV $(= 0.40 - (-0.16))$ for the former and 0.39 eV for the latter, is ascribed to an uncontrolled difference in the materials used in both of the measurements. The hole doping level of the MWA-treated sample was estimated to be 2.0 ~ $4.2\times10^{13}$ cm$^{-2}$ by the Hall measurement (Figure 5b). Unfortunately, the mobility substantially decreased to ~650 cm$^2$/Vs after MWA, in sharp contrast to the increase to 3000 cm$^2$/Vs after the hydrogen intercalation.[13, 14] A major cause for this degradation can be found in the onset of a little too high hole concentration ($p \approx 2.0 \sim 4.2 \times 10^{13}$ cm$^{-2}$) of the present material, which is compared to that ($p \approx 8.0\times10^{12}$ cm$^{-2}$) of the H-intercalated QFSEG.[14] Since the graphene's mobility is largely affected by the charged impurity scattering[45] and is thus inversely proportional to the carrier concentration, this increase in the hole concentration by a factor of 5 well accounts for the mobility degradation.

Despite the similar behavior with that of the conventional oxygen annealing methods, the processing time is dramatically reduced in MWA. Table 1 compares the process conditions for the oxygen intercalation between conventional oxygen annealing methods[20, 21, 46] and the present work together with the electrical properties of the resultant QFSEG. The QFSEG formed by the conventional method shows a high hole concentration of 1 ~ $4\times10^{13}$ cm$^2$ as well, accounting for the small carrier mobility of 60 ~ 790 cm$^2$/Vs. In the case of Kowalski *et al*,[46] the carrier mobility decreased from 300 cm$^2$/Vs to 60 cm$^2$/Vs after the oxygen intercalation. The electrical properties of the present QFSEG formed by MWA in air are quite similar or better to those of the conventional QFSEGs. In sharp contrast with these similarities in the properties, the treatment time has been reduced by more than one order of magnitude in MWA. The reason for this reduction in the processing time can be found in the more direct nature of the MWA's heating mechanism (Figure 6). In the conventional annealing method using a furnace system (Figure 6a), a tremendous amount of excess thermal energy must be supplied to the heater to heat the SiC substrate up to a desired temperature (~600 °C). Still, the substrate temperature can only be raised slowly (50 °C/min) and the annealing temperature must be maintained for 30 - 40 min to have a distinct effect. During annealing, oxygen atoms diffuse to the buffer layer and break the covalent bonding between C and Si atoms. Immediately after the breakage of the Si-C bonding, the oxygen atoms passivate the dangling bonds of the Si atoms, thus decoupling the ex-buffer layer from the SiC substrate. After the annealing at a high temperature, the sample in the furnace chamber is cooled down to a room temperature using typically 3 hours. The whole process thus needs a process time of as long as 5 hours or more. In MWA (Figure 6b), on the other hand, the substrate temperature can be raised to ~500 °C within 1 minute. This is due to the unique self-heating property of MWA, which eliminates the need for the heat transport from the heat source to the SiC substrate. The cooling rate is also very fast because of the absence of the residual heat in the ambient as well as the high thermal conductivity of the SiC substrate. These rapid heating/cooling rates may account for the dramatic decrease in the process time as well. It is reported for a Si-SiO$_2$ system that a rapid thermal annealing causes increase in the interface trap density, which was interpreted as to be caused by physical breakage of the Si-O bondings due to breakage of the different TEC values between Si and SiO$_2$.[47] In view of the negative TEC of graphene, it is natural to imagine that a similar mechanism is operative in our SiC/EG system during heating and cooling, and thereby greatly reduces the process time.

4. **Conclusions**

We proposed a novel and simple method for the formation of QFSEG using MWA in air. The results confirmed that a MWA on a mono-EG for just 2 min decouples the buffer layer from the SiC substrate and converts it into a bilayer QFSEG. The XPS results show that oxygen atoms passivate the Si dangling bonds caused by the breakage of the Si-C -bonding. The doping type changed from n-type ($n \approx 1 \times 10^{13}$ cm$^{-2}$) to p-type ($p \approx 3.0 \times 10^{13}$ cm$^{-2}$). Reflecting this increase in the carrier concentration, the carrier mobility decreased from ~1500 to ~650 cm$^2$/Vs, due primarily to the enhanced charged impurity scattering. This degradation will be overcome, for instance, by controlling the chamber ambient in the future. The substantial reduction of the processing time will surely contribute to the betterment of the productivity of the EG-based devices, and we believe that MWA method is a highly promising route towards nanoelectronic applications based on epitaxial graphene.

**Acknowledgement**

This work is supported by JSPS Grant-in-Aid for JSPS fellows (JP17J04947), and KAKENHI (JP16H00953).

**REFERENCES**

1) K. S. Novoselov, A. K. Geim, S. V. Morozov, D. Jiang, Y. Zhang, S. V. Dubonos, I. V. Grigorieva, and A. A. Firsov, Science **306**, 666 (2004).

2) A. K. Geim, Science **324**, 1530 (2009).

3) K. S. Novoselov, A. K. Geim, S. V. Morozov, D. Jiang, M. I. Katsnelson, I. V. Grigorieva, A. A. Dubonos, A. A. Firsov, Nature **438**, 197 (2005).

4) C. Berger, Z. Song, X. Li, X. Wu, N. Brown, C. Naud, D. Mayou, T. Li, J. Hass, A. N. Marchenkov, E. H. Conrad, P. N. First, W. A. de Heer, Science **312**, 1191(2006).

5) T. Ohta, A. Bostwick, J. L. McChesney, Th. Seyller, K. Horn, E. Rotenberg, Phys. Rev. Lett. **98**, 206802 (2007).

6) K. V. Emtsev, A. Bostwick, K. Horn, J. Jobst, G. L. Kellogg, L. Ley, J. L. McChesney, T. Ohta, S. A. Reshanov, J. Rohrl, E. Rotenberg, A. K. Schmid, D. Waldmann, H. B. Weber, T. Seyller, Nat. Mater. **8**, 203 (2009).

7) K. V. Emtsev, F. Speck, T. Seyller, L. Ley, Phys. Rev. B **77**, 155303 (2008).

8) W. Norimatsu, M. Kusunoki, Chem. Phys. Lett. **468**, 52 (2009).

9) C. Riedl, A. A. Zakharov, U. Starke, Appl. Phys. Lett. **93**, 033106 (2008).

10) J. Jobst, D. Waldmann, F. Speck, R. Hirner, D. K. Maude, T. Seyller, H. B. Weber, Phys. Rev. B **81**, 195434 (2010).

11) T. Ohta, A. Bostwick, T. Seyller, K. Horn, E. Rotenberg, Science **313**, 951 (2006).

12) C. Riedl, C. Coletti, T. Iwasaki, A. A. Zakharov, U. Starke, Phys. Rev. Lett. **103**, 246804 (2009).

13) F. Speck, J. Jobst, F. Fromm, M. Ostler, D. Waldmann, M. Hundhausen, H. B. Weber, Th. Seyller, Appl. Phys. Lett. **99**, 122106 (2011).

14) J. A. Robinson, M. Hollander, M. LaBella, K. A. Trumbull, R. Cavalero, D. W. Snyder, Nano Lett. **11**, 3875 (2011).

15) C. Virojanadara, S. Watcharinyanon, A. A. Zakharov, L. I. Johansson, Phys. Rev. B **82**, 205402 (2010).

16) K.V. Emtsev, A. A. Zakharov, C. Coletti, S. Forti, U. Starke, Phys. Rev. B **84**, 125423 (2011).

17) A. L. Walter, K. J. Jeon, A. Bostwick, F. Speck, M. Ostler, Th. Seyller, L. Moreschini, Y. S. Kim, Y. J. Chang, K. Horn, E. Rotenberg, Appl. Phys. Lett. **98**, 184102 (2011).

18) S. L. Wong, H. Huang, Y. Wang, L. Cao, D. Qi, I. Santoso, W. Chen, A. T. S. Wee, ACS Nano **5**, 7662 (2011).

19) I. Gierz, T. Suzuki, R. T. Weitz, D. S. Lee, B. Krauss, C. Riedl, U. Starke, H. Hochst, J. H. Smet, C. R. Ast, K. Kern, Phys. Rev. B **81**, 235408 (2010).

20) M. H. Oliveira Jr., T. Schumann, F. Fromm, R. Koch, M. Ostler, M. Ramsteiner, T. Seyller, J. M. J. Lopes, H. Riechert, Carbon **52**, 83 (2013).

21) M. Ostler, F. Fromm, R. J. Koch, P. Wehrfritz, F. Speck, H. Vita, S. Bottcher, K. Horn, T. Seyller, Carbon **70**, 258 (2014).

22) C. Mathieu, B. Lalmi, T. O. Mentes, E. Pallecchi, A. Locatelli, S. Latil, R. Belkhou, A. Ouerghi, Phys. Rev. B **86**, 035435 (2012).

23) J. Bao, W. Norimatsu, H. Iwata, K. Matsuda, T. Ito, M. Kusunoki, Phys. Rev. Lett. **117**, 205501 (2016).

24) D. Michael, P. Mingos, D. R. Baghurst, Chem. Soc. Rev. **20**, 1 (1991).

25) K. E. Haque, Int. J. Miner. Process. **57**, 1 (1999).

26) C. O. Kappe, Angew. Chem. Int. Ed. **43**, 6250 (2004).

27) L. F. Teng, P. T. Liu, Y. J. Lo, Y. J. Lee, Appl. Phys. Lett. **101**, 132901 (2012).

28) I. K. Lee, K. H. Lee, S. Lee, W. J. Cho, ACS Appl. Mater. Interfaces **6**, 22680 (2014).

29) T. L. Alford, D. C. Thompson, J. W. Mayer, N. D. Theodore, J. Appl. Phys. **106**, 114902 (2009).

30) H. M. A. Hassan, V. Abdelsayed, A. Khder, K. M. AbouZeid, J. Terner, M. S. El-Shall, S. I. Al-Resayes, A. A. El-Azhary, J. Mater. Chem. **19**, 3832 (2009).

31) V. Sridhar, J. H. Jeon, I. K. Oh, Carbon **48**, 2953 (2010).

32) Y. C. Kim, D. H. Cho, S. Ryu, C. G. Lee, Carbon **67**, 673 (2014).

33) K. H. Lee, J. W. Oh, J. G. Son, H. S. Kim, S. S. Lee, ACS Appl. Mater. Interfaces **6**, 6361 (2014).

34) D. E. Clark, D. C. Folz, R. L. Schuiz, Z. Fathi, A. D. Cozzi, MRS Bulletin **11**, 41(1993).

35) C. Riedl, U. Starke, J. Bernhardt, M. Franke, K. Heinz, Phys. Rev. B **76**, 245406 (2007).

36) J. G. Kim, E. J. Jung, Y. H. Kim, Y. Makarov, D. Choi, J. Ceram. Int. **40**, 3953 (2014).

37) D. S. Lee, C. Riedl, B. Krauss, K. V. Klitzing, U. Starke, J. H. Smet, Nano Lett. **8**, 4320 (2008).

38) A. C. Ferrari, J. C. Meyer, V. Scardaci, C. Casiraghi, M. Lazzeri, F. Mauri, S. Piscanec, D. Jiang, K. S. Novoselov, S. Roth, A. K. Geim, Phys. Rev. Lett. **97**, 187401 (2006).

39) T. M. G. Mohiuddin, A. Lombardo, R. R. Nair, A. Bonetti, G. Savini, R. Jalil, N. Bonini, D. M. Basko, C. Galiotis, N. Marzari, K. S. Novoselov, A. K. Geim, A. C. Ferrari, Phys. Rev. B **79**, 205433 (2009).

40) N. Ferralis, R. Maboudian, C. Carraro, Phys. Rev. Lett. **101**, 156801 (2008).

41) J. Rohrl, M. Hundhausen, K. V. Emtsev, Th. Seyller, R. Graupner, L. Ley, Appl. Phys. Lett. **92**, 201918 (2008).

42) A. H. C. Neto, F. Guinea, N. M. R. Peres, K. S. Novoselov, A. K. Geim, Rev. Mod. Phys. **81**, 109 (2009).

43) C. Riedl, C. Coletti, U. Starke, J. Phys. D: Appl. Phys. **43**, 374009 (2010).

44) J. Ristein, S. Mammadov, Th. Seyller, Phys. Rev. Lett. **108**, 246104 (2012).

45) J. H. Chen, C. Jang, S. Adam, M. S. Fuhrer, E. D. Williams, M. Ishigami, Nat. Phys. **4** 377 (2008).

46) G. Kowalski, M. Tokarczyk, P. Dabrowski, P. Ciepielewski, M. Mozdzonek, W. Strupinski, J. M. Baranowski, J. Appl. Phys. **117**, 105301 (2015).

47) P. K. Hurleya, A. Stesmans, V. V. Afanas'ev, B. J. O'Sullivan, E. O'Callaghan, J. Appl. Phys. **93**, 3971 (2003).

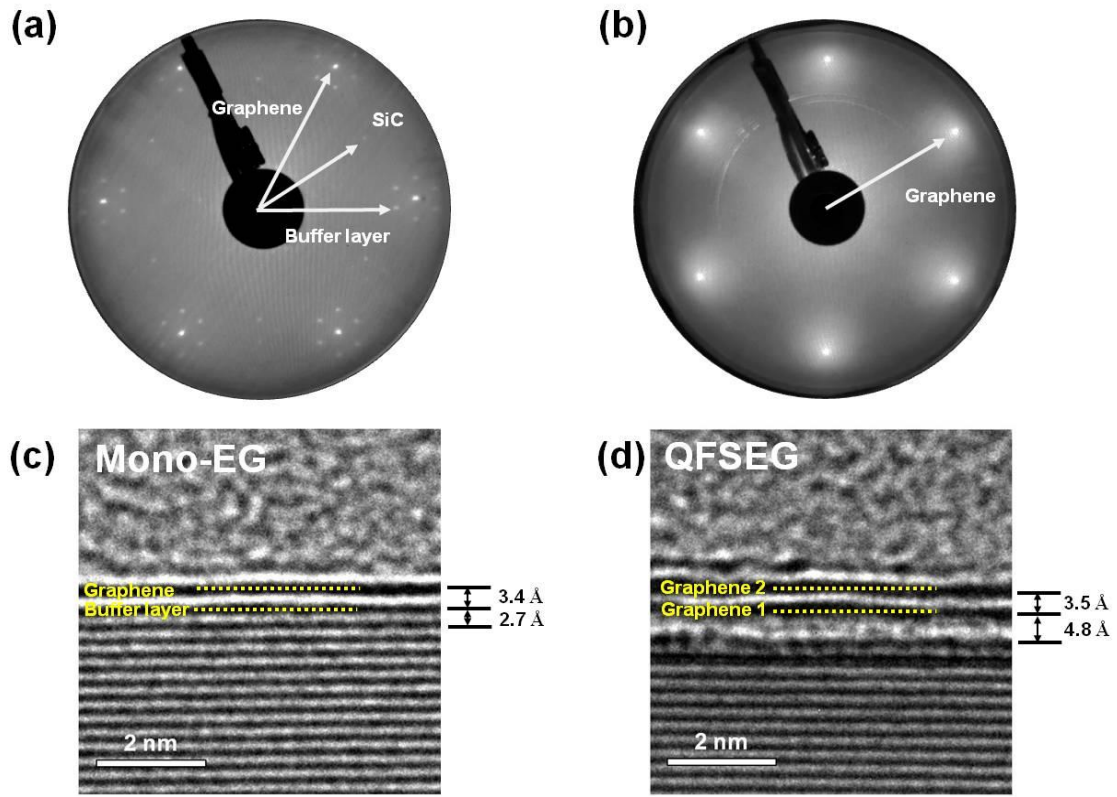

Figure 1. LEED patterns taken at 60 eV. Before MWA (a), a clear $(6\sqrt{3} \times 6\sqrt{3})R30º$ reconstruction pattern can be observed, which smears out (b) after MWA. HR-TEM images (c) before and (d) after MWA indicates conversion from (c) mono-EG to (d) QFSEG.

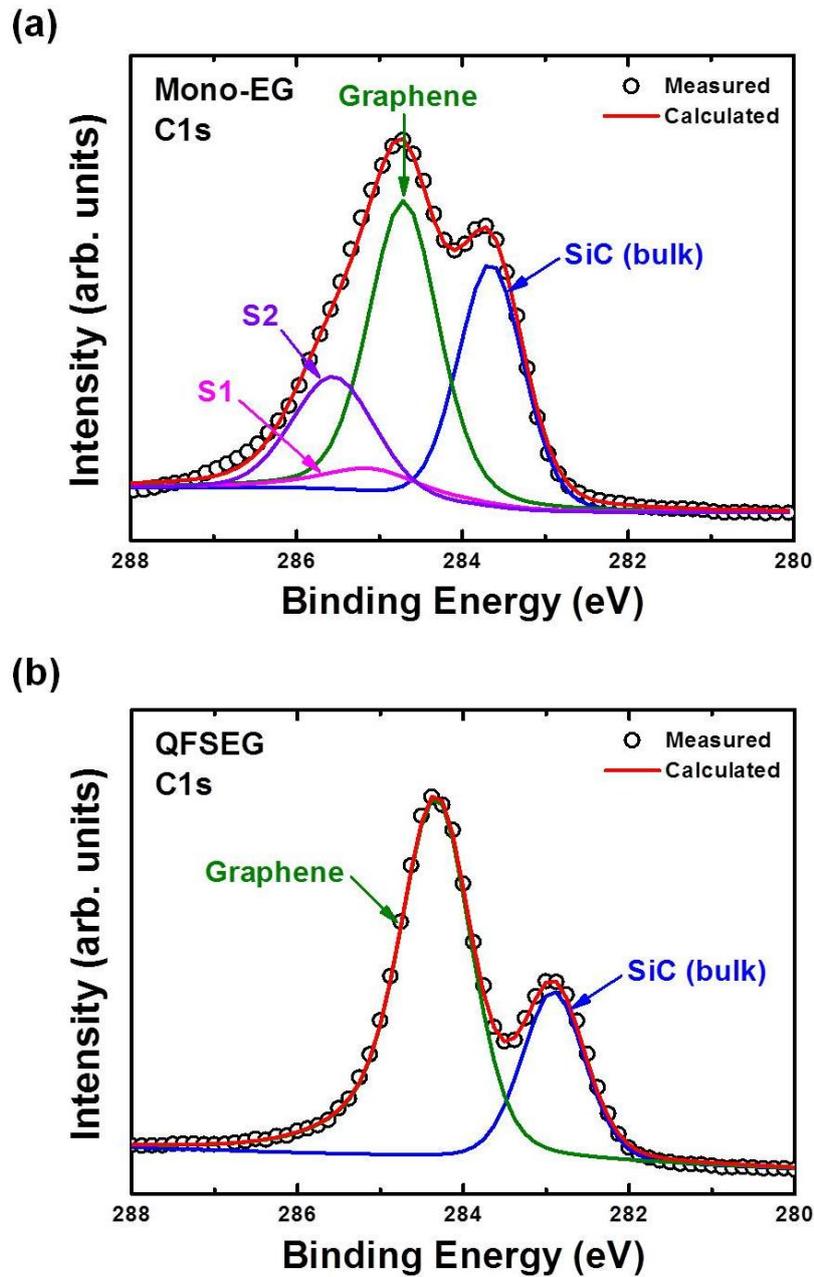

Figure 2. C1s core level spectrum (a) before and (b) after MWA. The experimental data are displayed in black open circles. The former is deconvoluted into components related with SiC, graphene and buffer layer (S1 and S2), showing presence of EG with the buffer layer. After MWA, the buffer-layer-related components disappear, indicating formation of QFSEG.

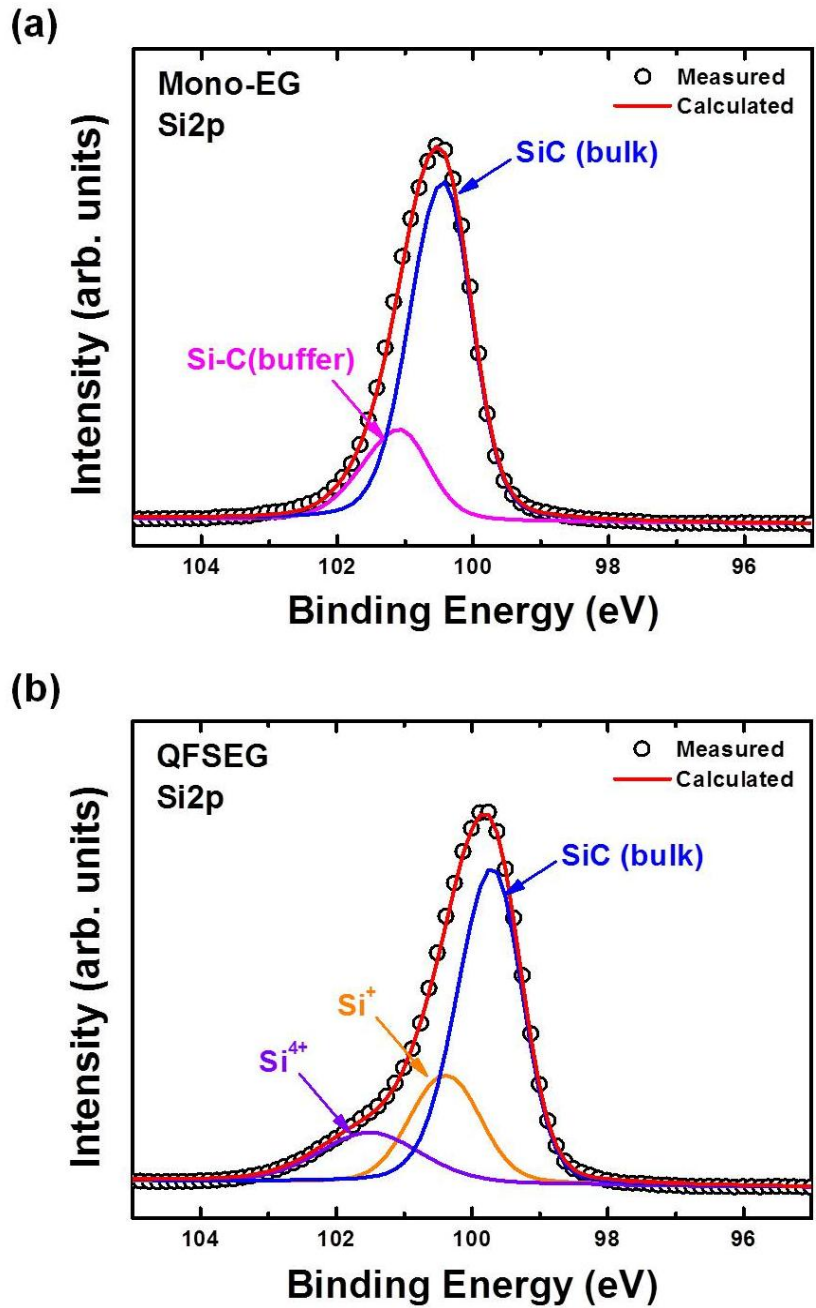

Figure 3. Si2p core level spectrum (a) before and (b) after MWA. In the latter, the Si-C bonding peak disappears and the oxide-related components ($Si^+$ and $Si^{4+}$) appear instead.

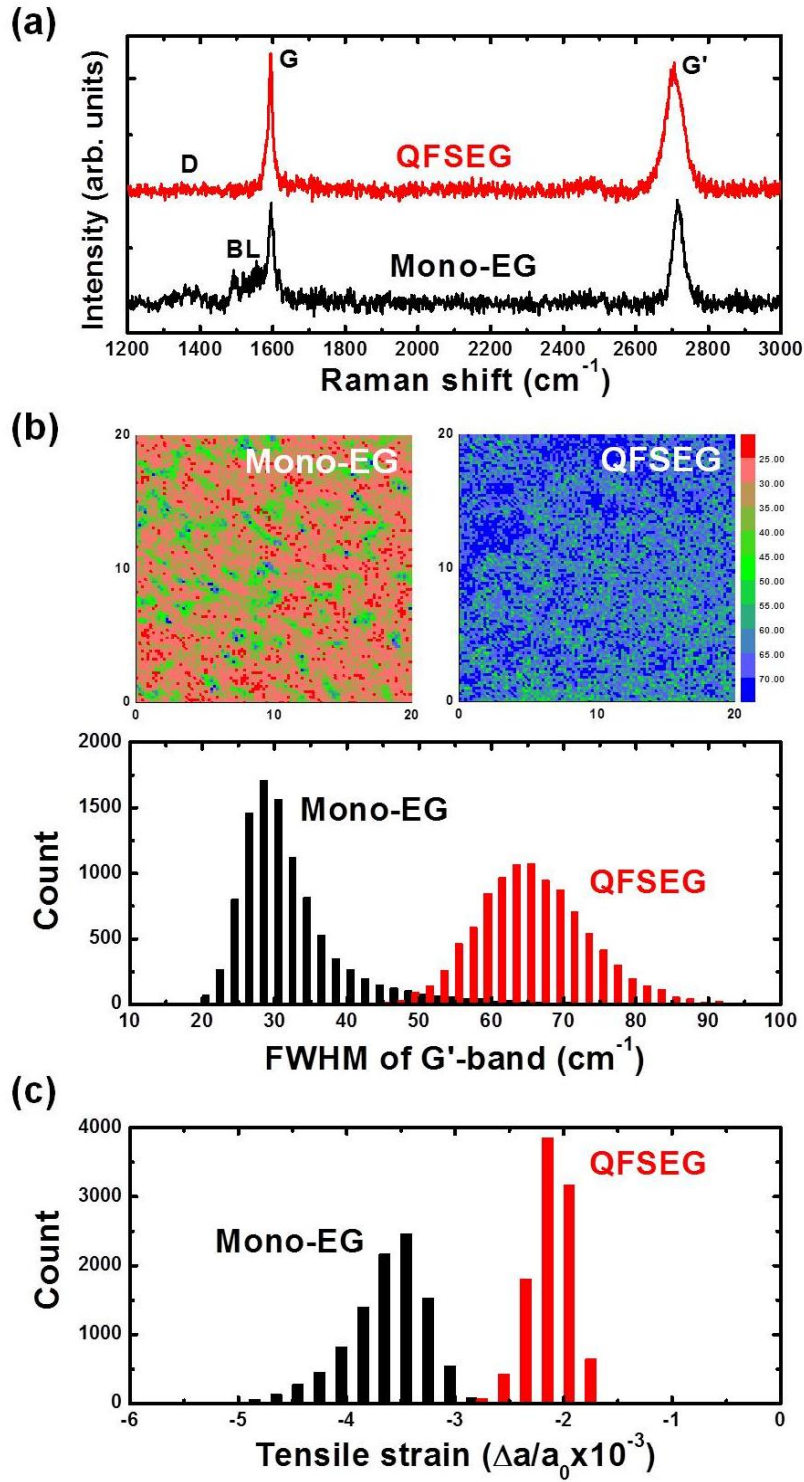

Figure 4. (a) Raman spectra before (black) and after (red) MWA. (b) Micro-Raman mapping of FWHM for G'-band (top) as well as its histogram (below) indicates conversion from monolayer graphene (left) to bilayer graphene (right) after MWA. (c) Comparison of strain, extracted from the G'-band shift, before (black) and after (red) MWA. The compressive strain present in for mono-EG is substantially relaxed in QFSEG.

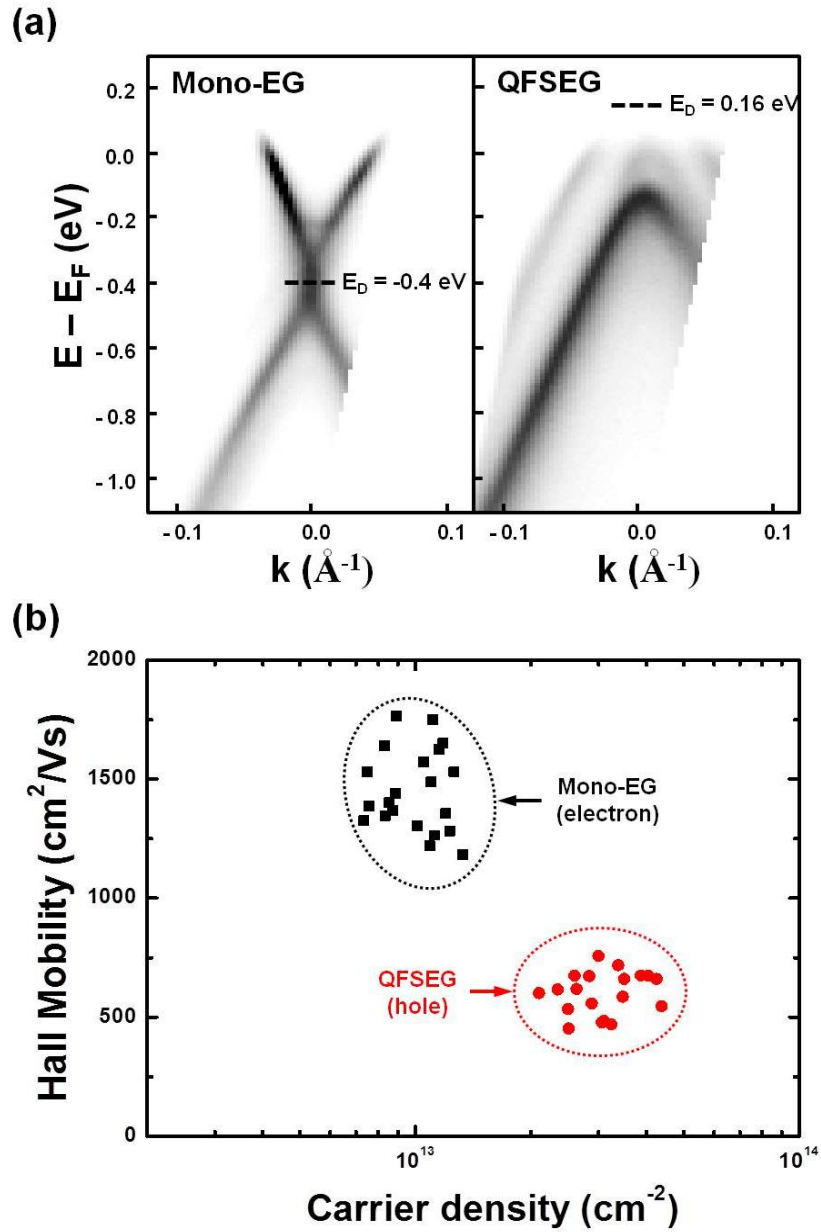

Figure 5. (a) Dispersion of the π-band at the K-point of the graphene Brillion zone for mono-EG and QFSEG. (b) Hall mobility v.s. the carrier density are compared between mono-EG (black) and QFSEG (red).

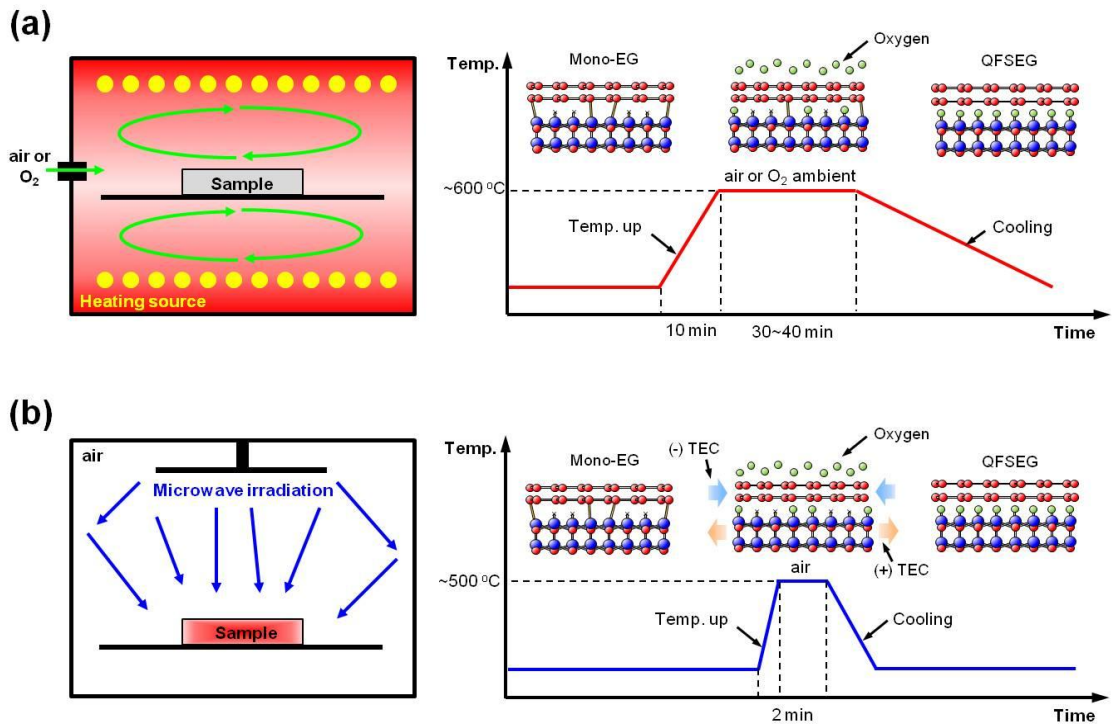

Figure 6. Comparison between (a) conventional furnace annealing and (b) MWA in air. While the heat is transferred from the heater to the sample via radiation and convection in (a), heat is generated inside the material in (b). Due to the more direct nature of the heating, MWA shows much higher heating/cooling rates (right diagrams).

| | Intercalation process | | | Carrier concentration | Carrier mobility | Ref. |
|---|---|---|---|---|---|---|
| Method | Temp. | Time | ambient | | | |
| Furnace | 600 °C | 40 min | Air | p ≈ 1×10$^{13}$ cm$^2$ | - | 20 |
| Furnace | 650 °C | 30 min | O$_2$ | p ≈ 2×10$^{13}$ cm$^2$ | 790 cm$^2$/Vs | 21 |
| Furnace | 600 °C | 30 min | Air | p ≈ 4×10$^{13}$ cm$^2$ | 60 cm$^2$/Vs | 46 |
| MWA | ~500 °C | 2 min | Air | p ≈ 4×10$^{13}$ cm$^2$ | 650 cm$^2$/Vs | |

Table 1. Process conditions and electrical properties of QFSEG compared between conventional furnace method and MWA.